\documentclass[12pt]{article}
\usepackage{amsmath}
\usepackage{graphicx,psfrag,epsf}
\usepackage{enumerate}
\usepackage{url} 

\newcommand{\blind}{0}

\usepackage{amsthm,amsfonts}
\usepackage{comment}
\usepackage[bf]{caption}
\usepackage{lscape}
\usepackage[backend=bibtex, style=nature, citestyle=authoryear]{biblatex}
\bibliography{bibliography}

\def\inprob{\stackrel{p}{\rightarrow}}
\def\indist{\rightsquigarrow}
\def\ind{\perp\!\!\!\perp}
\def\T{{ \mathrm{\scriptscriptstyle T} }}

\newcommand{\cov}{\text{cov}}
\newcommand{\Pb}{\mathbb{P}}
\newcommand{\Pn}{\mathbb{P}_n}

\newcommand{\E}{\mathbb{E}}
\newcommand{\R}{\mathbb{R}}

\newcommand{\bZ}{\mathbf{Z}}

\newcommand{\bX}{\mathbf{X}}
\newcommand{\bx}{\mathbf{x}}
\newcommand{\bY}{\mathbf{Y}}
\newcommand{\by}{\mathbf{y}}
\newcommand{\bV}{\mathbf{V}}
\newcommand{\bv}{\mathbf{v}}

\def\sd{\text{\textsc{sd}}}
\newcommand{\M}{\mathbb{M}}
\def\iqr{\text{\textsc{iqr}}}

\DeclareSymbolFont{bbold}{U}{bbold}{m}{n}
\DeclareSymbolFontAlphabet{\mathbbold}{bbold}
\newcommand{\one}{\mathbbold{1}}

\theoremstyle{remark}
\newtheorem{assumption}{Assumption}

\begin{document}

\def\spacingset#1{\renewcommand{\baselinestretch}%
{#1}\small\normalsize} \spacingset{1}


\if0\blind
{
  \title{ \vspace*{-.5in} \bf Estimating scaled treatment effects \\ with multiple outcomes}
  \author{Edward H. Kennedy
    \thanks{Edward Kennedy is Assistant Professor in the Department of Statistics, Carnegie Mellon University, Pittsburgh, PA 15213 (e-mail: edward@stat.cmu.edu). Edward Kennedy gratefully acknowledges support from NIH grant R01-DK090385, and Shreya Kangovi and Nandita Mitra from PCORI grant AD-1310-0792. The authors thank Emin Tahirovic for helpful discussions and programming support on an earlier version of this manuscript.}\hspace{.2cm}
    \\
    Department of Statistics, Carnegie Mellon University \\ \\
    Shreya Kangovi \\
    Division of General Internal Medicine, University of Pennsylvania \\ \\
    Nandita Mitra \\
    Department of Biostatistics \& Epidemiology, University of Pennsylvania \\ \\ }
  \date{}
  \maketitle
  \setcounter{page}{0}
  \thispagestyle{empty}
} \fi

\if1\blind
{
  \bigskip
  \bigskip
  \bigskip
  \begin{center}
    {\LARGE\bf Estimating scaled treatment effects \\ with multiple outcomes}
\end{center}
  \medskip
} \fi

\bigskip
\vspace*{-.45in}
\begin{abstract}
In classical study designs, the aim is often to learn about the effects of a treatment or intervention on a single outcome; in many modern studies, however, data on multiple outcomes are collected and it is of interest to explore effects on multiple outcomes simultaneously. Such designs can be particularly useful in patient-centered research, where different outcomes might be more or less important to different patients.  In this paper we propose scaled effect measures (via potential outcomes) that translate effects on multiple outcomes to a common scale, using mean-variance and median-interquartile-range -based standardizations. We present efficient, nonparametric, doubly robust methods for estimating these scaled effects (and weighted average summary measures), and for testing the null hypothesis that treatment affects all outcomes equally. We also discuss methods for exploring how treatment effects depend on covariates (i.e., effect modification). In addition to describing efficiency theory for our estimands and the asymptotic behavior of our estimators, we illustrate the methods in a simulation study and a data analysis. Importantly, and in contrast to much of the literature concerning effects on multiple outcomes, our methods are nonparametric and can be used not only in randomized trials to yield increased efficiency, but also in observational studies with high-dimensional covariates to reduce confounding bias.
\end{abstract}

\noindent%
\vfill

\thispagestyle{empty}

\newpage

\spacingset{1.45} 
\spacingset{1}

\section{Introduction}
\label{s:intro}

In classical study designs, the aim is often to learn about the effects of a treatment or intervention on a single outcome; in many modern studies, however, data on multiple outcomes are collected and it is of interest to explore effects on multiple outcomes simultaneously.  Such designs are particularly important in patient-centered research, for example, where different outcomes might be more or less important to different patients \autocite{kangovi2014patient, kangovi2017randomized}, and more generally in prioritizing public health recommendations \autocite{vanderweele2017outcome}. 

There has been varied and relatively extensive discussion in the literature over the past few decades about estimating treatment effects on multiple outcomes \autocite{obrien1984procedures,pocock1987analysis,sammel1999multivariate,freemantle2003composite,thurston2009bayesian,teixeira2011statistical,yoon2011alternative}, including estimating scaled effects \autocite{lin2000scaled,roy2003scaled}, which is a major focus of this paper. However, most of the aforementioned work requires strong parametric assumptions and is geared towards randomized trials rather than observational studies, which can require adjustment for high-dimensional confounders. In contrast, we consider nonparametric doubly robust methods for estimation and hypothesis testing of scaled treatment effects on multiple outcomes. In particular we translate effects to a common scale with mean-variance and median-interquartile-range -based standardizations, which are constructed within an explicitly causal potential outcomes framework. Our work is a response to recent proposals by \textcite{vanderweele2017outcome} and others to spend more effort exploring effects of interventions on multiple outcomes simultaneously, rather than using the classical one-outcome-at-a-time approach.  Importantly our work is designed to accommodate modern studies that include complex covariate information, which can be leveraged for efficiency gains or to reduce confounding bias (or both). 

The setup of the paper is as follows. In Sections 3.1 and 3.2 we present efficient doubly robust methods for estimating our proposed scaled effects (along with weighted average summary measures in Section 3.5), as well as methods for testing the null hypothesis that treatment affects all outcomes equally in Section 3.3. We also discuss approaches for exploring how treatment effects vary with covariates (i.e., effect modification) in Section 3.4. In addition to describing efficiency theory for our estimands and the asymptotic behavior of our estimators, in Section 4 we illustrate the methods in a simulation study and in Section 5 we apply them to a recently conducted trial evaluating the effect of community health workers on various health outcomes in a low income population. 

\section{Setup}
\label{s:setup}

\subsection{Data \& Notation}

We suppose we observe an independent and identically distributed sample $(\bZ_1,...,\bZ_n)$, where each observation $\bZ=(\bX,A,\bY)$ consists of a vector of $p$ covariates $\bX=(X_1,...,X_p)$, a binary treatment $A$, and a vector of $K$ outcomes $\bY=(Y_1,..,Y_K)$. We characterize treatment effects using potential outcome notation \autocite{rubin1974estimating}, letting $\bY^a=(Y_1^a, ..., Y_K^a)$ denote the outcome vector that would have been observed under treatment level $a$.

We use $\Pb$ to denote the distribution of $\bZ=(\bX,A,\bY)$, and write expectations under $\Pb$ with usual $\E$ operator notation. For a generic random variable $U$ we define standard deviations as usual with $\sd(U)=\sqrt{\E(U^2)-\E(U)^2}$. We use $\Pn$ to denote the empirical measure so that sample averages can be written as $\frac{1}{n} \sum_i f(\bZ_i) = \Pn\{f(\bZ)\}$. Finally we use the following notation to simplify the presentation:
\begin{align*}
\pi(a \mid \bx) &= \Pb(A=a \mid \bX=\bx) \\
\mu_k(\bx,a) &= \E(Y_k \mid \bX=\bx, A=a) \\
\eta_k(\bx,a) &= \E(Y_k^2 \mid \bX=\bx, A=a) .
\end{align*}

\subsection{Identification}

Throughout this paper we consider estimating quantities defined in terms of the distributions of potential outcomes $\bY^a$ for $a=0,1$. Since potential outcomes are not observed directly, we need identifying assumptions to express estimands of interest in terms of the estimable observed data distribution. We consider the usual ignorability or ``no unmeasured confounding'' setting, in which the following assumptions hold for $a=0,1$: 

\medskip

\begin{assumption} \label{ass1}
Consistency: $A=a$ implies $\bY=\bY^a$.
\end{assumption}
\begin{assumption} \label{ass2}
Positivity: $\Pb\{ 0< \pi(a \mid \bX) < 1\}=1$ for all $a$.
\end{assumption}
\begin{assumption} \label{ass3}
Exchangeability: $A \ind \bY^a \mid \bX$.
\end{assumption} 

\medskip

Assumptions \ref{ass1}--\ref{ass3} can hold by design in a randomized trial, since treatment $A$ is under the control of investigators. However in observational studies these assumptions can be violated, and are generally untestable (apart from positivity). Consistency means potential outcomes are defined uniquely by subjects' own treatment levels (this can be violated in, for example, vaccine studies). Positivity means treatment is not assigned deterministically for any subjects, regardless of covariates. Exchangeability means treatment is as good as randomized (within covariate strata) since it is unrelated to potential outcomes once we condition on covariates. Exchangeability requires either external randomization of treatment, or else the collection of sufficiently many relevant covariates.  

Assumptions \ref{ass1}--\ref{ass3} have been discussed at length elsewhere, and it is well-known that they imply 
$$ \Pb(\bY^a \leq \by \mid \bX) = \Pb(\bY \leq \by \mid \bX, A=a),$$
i.e., the conditional distribution of potential outcomes under $A=a$ (given covariates) equals the conditional distribution of observed outcomes (given covariates) among those for whom  $A=a$ observationally. For example this fact also implies that $\E(\bY^a) = \E\{ \E(\bY \mid \bX,A=a) \}$ and similarly for other marginal quantities.

\section{Methodology}
\label{sec:method}

In this section we present scaled treatment effect parameters and estimators, discuss how to test for differential effects across multiple outcomes, give extensions for exploring how treatment effects vary with covariates, and finally present  weighted average measures that can provide a scalar summary of multivariate effects.

\subsection{Scaled Average Effects}
\label{sub:avg}

We start by presenting a scaled treatment effect parameter (using mean-variance standardization), discuss corresponding semiparametric efficiency theory,  give doubly robust and locally efficient estimators, and describe asymptotic properties.

As noted for example by \textcite{lin2000scaled,roy2003scaled}, usual outcome-specific effects cannot be compared directly in studies with multiple outcomes measured on different scales. For example, average differences of the form $\E(Y_k^1 - Y_k^0)$ will generally have different and non-comparable units (e.g., kilograms for $k=1$ and millimeters of mercury for $k=2$). Thus, to generate a unitless measure of effect, we propose a simple standardization by the standard deviation of outcomes $Y_k^0$ under control. Note that this is most useful in settings where $A=0$ represents a meaningful control group (e.g., standard of care), rather than an alternative and potentially comparable treatment option. 

Specifically we characterize effects on outcome $k$ with the scaled effect measure
\begin{equation} \label{eq:psi}
\psi_k = \frac{ \E(Y_k^1 - Y_k^0) }{ \sd(Y_k^0) } .
\end{equation}
This effect measure captures the mean difference in outcomes under treatment versus control, expressed in terms of the standard deviation under control. Thus $\psi_k=1$ indicates that treatment increases outcomes by one standard deviation, on average, of what they would have been under control; similarly $\psi_k=2$ means treatment increases outcomes by two standard deviations on average, and $\psi_k=-1$ means treatment decreases outcomes by one standard deviation. Scaled effect measures have played an important role in studies with multiple outcomes \autocite{lin2000scaled,roy2003scaled}, however so far they have only been proposed within the context of parametric models. Our work can thus be viewed as a nonparametric extension, which also admits doubly robust estimators. 

We can use results from semiparametric theory \autocite{bickel1993efficient, van2003unified, tsiatis2006semiparametric} to construct optimal estimators for $\psi_k$ under minimal assumptions about the distribution of the data $\Pb$. We refer to \textcite{kennedy2016semiparametric} for a review. First define, for $a=0,1$,
\begin{align}
\phi_{ak}(\bZ;\pi,\mu) &= \frac{\one(A=a) }{\pi(a \mid \bX)} \Big\{Y_k - \mu_k(\bX,a) \Big\} + \mu_k(\bX,a) \\
\phi_{2k}(\bZ;\pi,\eta) &= \frac{\one(A=0)}{\pi(0 \mid \bX)}  \Big\{Y_k^2 - \eta_k(\bX,0) \Big\} + \eta_k(\bX,0) .
\end{align}
as components of the efficient influence functions for $\E(Y_k^a)$ and $\E\{(Y_k^0)^2\}$, respectively. Then, given estimators $(\hat\pi,\hat\mu,\hat\eta)$ of the nuisance functions, the estimator
\begin{equation} \label{eq:psihat}
\hat\psi_k = \frac{ \Pn\Big\{ \phi_{1k}(\bZ;\hat\pi,\hat\mu) - \phi_{0k}(\bZ;\hat\pi,\hat\mu) \Big\} }{ \sqrt{\Pn\Big\{ \phi_{2k}(\bZ;\hat\pi,\hat\eta) \Big\} - \Big[\Pn\Big\{ \phi_{0k}(\bZ;\hat\pi,\hat\mu)\Big\} \Big]^2 } }
\end{equation}
is doubly robust and locally efficient, under nonparametric models as well as models that put some (e.g., parametric) restrictions on the treatment mechanism $\pi$. We will now discuss these properties in more detail, proofs of which are given in the Supplementary Materials.

Double robustness is a very important property that has been discussed in detail before \autocite{robins2001inference,bang2005doubly}. One important consequence of double robustness is that analysts have two chances at obtaining a consistent estimator. For example, in our case, the estimator $\hat\psi_k$ is consistent for its target $\psi_k$ as long as either of the nuisance estimators $\hat\pi$ or $(\hat\mu,\hat\eta)$ are consistent, even if one of $\hat\pi$ or $(\hat\mu,\hat\eta)$ is misspecified. In particular, this means consistency of $\hat\psi_k$ is guaranteed in a randomized trial, since there $\pi$ is known and thus can be estimated consistently under no assumptions. We prove that our estimator is doubly robust in Section 1 of the Supplementary Materials.

Another crucially important property of doubly robust estimators is that they can attain fast parametric $\sqrt{n}$ rates of convergence even after machine learning-based covariate adjustment \autocite{van2011targeted}. This is not the case for most standard plug-in estimators, which typically inherit slower-than-$\sqrt{n}$ convergence rates from their nuisance estimators \autocite{van2014higher}. In contrast, our proposed estimator $\hat\psi_k$ will be $\sqrt{n}$-consistent and asymptotically normal even if the nuisance functions $(\hat\pi,\hat\mu,\hat\eta)$ are estimated flexibly with nonparametric or machine learning methods, as long as these nuisance estimators converge at faster than $n^{1/4}$ rates (and under some empirical process conditions, which can be avoided with sample splitting). 

In particular, under the above conditions and other standard regularity conditions given in the Supplementary Materials, $\hat\psi_k$ is asymptotically normal,
\begin{equation} \label{eq:asym_norm}
\sqrt{n}(\hat\psi_k - \psi_k) \indist N(0,\sigma_k^2)  ,
\end{equation}
with asymptotic variance $\sigma_k^2$ equal to the variance of the efficient influence function, which is given by $\varphi_k(\bZ;\pi,\mu,\eta)$ defined as
\begin{align} \label{eq:eif}
& \frac{\phi_{1k}(\bZ;\pi,\mu) - \phi_{0k}(\bZ;\pi,\mu)}{\sd(Y_k^0)}  - \psi_k \left[ \frac{ \phi_{2k}(\bZ;\pi,\eta) + \E\{(Y_k^0)^2\} - 2 \E(Y_k^0) \phi_{0k}(\bZ;\pi,\mu) }{2 \ \sd(Y_k^0)^2} \right]  .
\end{align}
Thus Wald-type confidence intervals for $\psi_k$ can be constructed by estimating the asymptotic variance $\sigma_k^2$ with the empirical variance of the estimated efficient influence function values (obtained by replacing unknown quantities in \eqref{eq:eif} with estimates). The asymptotic normality result \eqref{eq:asym_norm} is proved in Section 2 of the Supplementary Materials, using empirical process theory \autocite{van1996weak,van2000asymptotic} to allow for flexible nonparametric estimation of $(\hat\pi,\hat\mu,\hat\eta)$. We prove that \eqref{eq:eif} is in fact the efficient influence function in Section 3 of the Supplementary Materials, which (given the asymptotic normality result) implies that $\hat\psi_k$ is locally efficient. In particular $\hat\psi_k$ is locally semiparametric efficient under models that put at most some restrictions on the treatment mechanism (including for example nonparametric models, as well as models in which the treatment mechanism is known). 

Finally, in Section 6 of the Supplementary Materials we give easily implementable R code for computing $\hat\psi_k$ based on estimating equations and targeted maximum likelihood (TMLE) \autocite{van2006targeted,van2011targeted}. TMLE uses specially constructed nuisance estimates $(\hat\mu^*,\hat\eta^*)$ that insure that resulting estimators of the numerator and denominator of \eqref{eq:psihat} respect the bounds of the parameter space (e.g., the numerator must lie between $[-1,1]$ when $Y_k \in [0,1]$), which can improve finite-sample properties and lessen the impact of extreme propensity scores. The provided code also calculates confidence intervals based on the approach described in the previous paragraph.

\subsection{Scaled Quantile Effects}
\label{sub:quantile}

In some cases, the mean and variance are not useful measures of centrality and spread (e.g., for distributions that are highly skewed), in which case the standardization given in \eqref{eq:psi} may not be most appropriate. An alternative quantile-based standardization that is immune to such concerns is given by
\begin{equation} \label{eq:psiq}
\psi_k^q = \frac{ \M(Y_k^1) - \M(Y_k^0) }{ \iqr(Y_k^0) } ,
\end{equation}
where for arbitrary random variable $U$ with distribution function $F(u) = P(U \leq u)$, we define $\xi(q) = \inf\{u: q \leq F(u) \}$ as the $q$-th quantile, and let $\M(U)=\xi(0.50)$ denote the median and $\iqr(U)=\xi(0.75)-\xi(0.25)$ denote the interquartile range. Therefore $\psi_k^q$ captures the difference in median outcomes under treatment versus control, expressed in terms of the interquartile range under control. For example, $\psi_k^q=0.5$ means treatment increases the median outcome, by half the interquartile range under control.

As for the scaled average effect in \eqref{eq:psi}, a doubly robust and locally efficient estimator for the quantile effect $\psi_k^q$ is given by
\begin{equation}
\hat\psi_k^q = \frac{ \hat{F}_{1k}^{-1}(0.50) -  \hat{F}_{0k}^{-1}(0.50) }{ \hat{F}_{0k}^{-1}(0.75) -  \hat{F}_{0k}^{-1}(0.25) }
\end{equation}
where $\hat{F}_{ak}^{-1}(\cdot)$ is the inverse of $\hat{F}_{ak}(y) = \Pn\{ \phi_{ak}^{(y)}(\bZ;\hat\pi,\hat\nu) \}$ (up to order $o_\Pb(1/\sqrt{n})$ if exact solutions cannot be found), for
\begin{align}
\phi_{ak}^{(y)}(\bZ;\pi,\nu) =  \frac{\one(A=a) }{\pi(a \mid \bX)} & \Big\{\one(Y_k \leq y) - \nu_k(y \mid \bX,a) \Big\}  + \nu_k(y \mid \bX,a) 
\end{align}
where $\nu_k(y \mid \bX,a) = \Pb(Y_k \leq y \mid \bX=\bx,A=a)$. We give the efficient influence function for $\psi_k^q$ (as well as conditions under which this is the influence function for $\hat\psi_k^q$) in Section 4 of the Supplementary Materials; results for unscaled effects were developed by \textcite{diaz2015efficient}.

\subsection{Hypothesis Testing}
\label{sub:test}

In what follows, we suppose for concreteness that the mean-variance standardization in \eqref{eq:psi} is appropriate for all $K$ outcomes. However, all results can be equally extended to the quantile-based standardization given in \eqref{eq:psiq} (if mean-variance standardization is only appropriate for some covariates, we suggest using quantile-based standardization for all).

As mentioned in the Introduction, in studies with multiple outcomes it is often of interest to assess whether any outcomes are differentially affected by treatment, and if so, which outcomes. Additional motivation is given in Section 5, as well as by \textcite{lin2000scaled,roy2003scaled}. This goal can be accomplished by testing a hypothesis of the form
\begin{equation} \label{eq:hyp}
H_0: \psi_1 = \psi_2 = ... = \psi_K
\end{equation}
which says that all scaled treatment effects are equal. Recall that this is a meaningful hypothesis even if outcomes are measured on different scales, due to the fact that the scaled effects are unitless after dividing by standard deviations under control.

A doubly robust test of the hypothesis in \eqref{eq:hyp} can be constructed based on the asymptotic distribution of 
\begin{equation}
T_n = n (\mathbf{C} \boldsymbol{\hat\psi})^\T (\mathbf{C} \boldsymbol{\hat\Sigma} \mathbf{C}^\T)^{-1}  (\mathbf{C} \boldsymbol{\hat\psi}),
\end{equation}
where $\mathbf{C}$ is a $(K-1) \times K$ banded matrix with elements 
$ C_{ij} = \one(i=j) - \one(i=j-1)$, 
 $\boldsymbol{\hat\psi}=(\hat\psi_1,...,\hat\psi_K)^\T$, and $\boldsymbol{\hat\Sigma}$ is an estimator of the asymptotic variance $\boldsymbol\Sigma$ of $\boldsymbol{\hat\psi}$ (estimation of which is discussed in the next paragraph). Specifically, under conditions given in Sections 2 and 5 of the Supplementary Materials, we have that 
\begin{equation}
T_n  \indist \chi^2_{K-1}
\end{equation}
under the null hypothesis $H_0$ of homogeneous effects given in \eqref{eq:hyp}. Therefore an asymptotic $p$-value for testing $H_0$ is given by $P(\chi^2_{K-1} \geq t_n)$, where $t_n$ is the observed value of $T_n$ in the sample (and $\chi^2_{K-1}$ is a chi-squared random variable with $K-1$ degrees of freedom). 

If the conditions given in Sections 2 and 5 of the Supplementary Materials hold, then a closed-form estimator for $\boldsymbol\Sigma$ can be obtained by replacing unknown quantities in \eqref{eq:eif} with estimates and computing the empirical covariance
$$ \boldsymbol{\hat\Sigma} = \Pn \Big\{ \boldsymbol{\varphi}(\bZ;\hat\pi,\hat\mu,\hat\eta)^{\otimes 2} \Big\} $$
where $\boldsymbol{\varphi}=(\varphi_1,...,\varphi_K)^\T$ is a vector of the stacked influence functions from \eqref{eq:eif} for $k=1,...,K$, and $\mathbf{u}^{\otimes 2} = \mathbf{u} \mathbf{u}^\T$ for any vector $\mathbf{u}$. Note that the estimated influence functions used to construct the estimator $\boldsymbol{\hat\Sigma}$ will also depend on estimates of $\widehat\E\{(Y_k^0)^2\} = \Pn\{ \phi_{2k}(\bZ;\hat\pi,\hat\eta) \}$, $\widehat\E(Y_k^0) =  \Pn\{ \phi_{0k}(\bZ;\hat\pi,\hat\mu)\}$, and $\widehat\sd(Y_k^0)^2 = \hat\E\{(Y_k^0)^2\} - \{\hat\E(Y_k^0)\}^2$. 

The conditions for the above estimator to be valid require that the product of convergence rates for $\hat\pi$ and $(\hat\mu,\hat\eta)$ is faster than $\sqrt{n}$, for example if $\hat\pi$ is estimated with a correct parametric model or known and $(\hat\mu,\hat\eta)$ is merely consistent (so that the product is $O_\Pb(1/\sqrt{n})o_\Pb(1)=o_\Pb(1/\sqrt{n})$), or if $(\hat\pi,\hat\mu,\hat\eta)$ are all consistent and converge at faster than $n^{1/4}$ rates (so that the product is $o_\Pb(n^{-1/4})o_\Pb(n^{-1/4})=o_\Pb(1/\sqrt{n})$). There is one setting where the above approach is valid even if this condition on the product of convergence rates does not hold. Specifically, if $\hat\pi$ is estimated with a correct parametric model, then even if $(\hat\mu,\hat\eta)$ is misspecified and the estimator $\boldsymbol{\hat\Sigma}$ is thus inconsistent, the above approach gives conservative $p$-values and is still valid. This is a result of the fact that estimating the propensity score $\pi$ when it is actually known cannot decrease (and will generally increase) efficiency \autocite{tsiatis2006semiparametric}; thus $\boldsymbol{\hat\Sigma} \geq \boldsymbol{\Sigma}$ in the sense that $\boldsymbol{\hat\Sigma} - \boldsymbol{\Sigma}$ is a positive definite matrix.

Alternatively the bootstrap can also be used to construct the estimator $\boldsymbol{\hat\Sigma}$; such an approach would be valid as long as $\boldsymbol{\hat\psi}$ is asymptotically linear, which is a weaker condition than requiring the product of convergence rates to be $o_\Pb(1/\sqrt{n})$. (For example, asymptotic linearity would hold in the scenarios discussed above, as well as if $(\hat\pi,\hat\mu,\hat\eta)$ were estimated with parametric models and it was only assumed that either $\hat\pi$ or $(\hat\mu,\hat\eta)$ were correctly modeled.) In practice the bootstrap might be preferred for computing $\boldsymbol{\hat\Sigma}$ since it depends on weaker assumptions, although it is more computationally expensive.

To test which outcomes are differentially affected by treatment, we can test the pairwise hypotheses $H_{jk}: \psi_j=\psi_k$. To control the family-wise error rate (which might be reasonable if $K$ is not too large) a simple Bonferroni correction could be used. Alternatively, to control the false discovery rate (which might be preferable if $K$ is large), the Benjamini-Hochberg procedure could be used instead.

\subsection{Effect Modification}
\label{sub:effectmod}

Often it is of interest to go beyond marginal effects like $\E(Y_k^1 - Y_k^0)$ or the scaled version in \eqref{eq:psi}, and further assess how treatment effects vary with covariates. This can be useful for exploring the mechanism by which treatment actually works, as well as for learning how to tailor treatment decisions to individual patient characteristics (since treatments may only work for some patients, or may be harmful for some and beneficial for others). 

A natural extension of the standard effect parameter in \eqref{eq:psi} that allows for assessing such effect modification is given by
\begin{equation}
\gamma_k(\bv) = \frac{ \E(Y_k^1 - Y_k^0 \mid \bV=\bv) }{ \sd(Y_k^0 \mid \bV=\bv) } ,
\end{equation}
where $\bV \subseteq \bX$ is a subset of the full covariate set that only includes the variables for which effect modification is of interest. Similar to prior subsections, this effect measures the mean difference in outcomes under treatment versus control for those with covariates $\bV=\bv$, in terms of the standard deviation under control for this same group.

When $\bV$ only contains a modest number of discrete variables, the estimator in \eqref{eq:psihat} can be easily modified to estimate $\gamma_k(\bv)$ with
\begin{equation} \label{eq:gammahat}
\frac{ \Pn\Big\{ \phi^{(\bv)}_{1k}(\bZ;\hat\pi,\hat\mu) - \phi^{(\bv)}_{0k}(\bZ;\hat\pi,\hat\mu) \Big\} }{ \sqrt{\Pn\Big\{ \phi^{(\bv)}_{2k}(\bZ;\hat\pi,\hat\eta) \Big\} - \Big[\Pn\Big\{ \phi^{(\bv)}_{0k}(\bZ;\hat\pi,\hat\mu)\Big\} \Big]^2 } }
\end{equation}
where the functions $ \phi^{(\bv)}_{ak}(\bZ;\pi,\mu) =  \phi_{ak}(\bZ;\pi,\mu) \one(\bV=\bv) / \Pn\{\one(\bV=\bv)\}$ and $ \phi^{(\bv)}_{2k}(\bZ;\pi,\mu) =  \phi_{2k}(\bZ;\pi,\mu) \one(\bV=\bv) /\Pn\{\one(\bV=\bv)\}$ are $\bv$-specific versions of the influence functions from previous sections, so that the averages in \eqref{eq:gammahat} are just over those units with $\bV=\bv$. The influence function for the estimator in \eqref{eq:gammahat} is given in the next subsection.

When $\bV$ contains a continuous variable or many discrete variables, the above approach will not be feasible since the cells $\bV=\bv$ will be very small or empty. There are a few options for such cases. First, one could estimate $\E(Y_k^a \mid \bV)$ and $\E\{(Y_k^a)^2 \mid \bV\}$ by regressing $\phi_{ak}(\bZ;\hat\pi,\hat\mu)$ and $\phi_{2k}(\bZ;\hat\pi,\hat\eta)$ on $\bV$ using any preferred methods, such as parametric regression modeling or flexible machine learning, and then construct
$$ \hat\gamma_k(\bv) = \frac{ \widehat\E(Y_k^1 \mid \bV=\bv) - \widehat\E(Y_k^0 \mid \bV=\bv) }{ \sqrt{ \widehat\E\{(Y_k^0)^2 \mid \bV=\bv \} - \{ \widehat\E(Y_k^0 \mid \bV=\bv) \}^2 } }  $$
based on these regressions. A potential disadvantage of this approach is that $\hat\gamma_k(\bv)$ will in general not follow an interpretable parametric model, even if the models for the components $\E(Y_k^a \mid \bV)$ and $\E\{(Y_k^a)^2 \mid \bV\}$ do; e.g., the ratio of two linear models is not itself linear. 

To combat this problem, one could regress the predicted values from the above approach onto a parametric model $\gamma_k(\bv;\boldsymbol\theta)$ for some $\boldsymbol\theta \in \R^q$ (for example, the linear model $\gamma_k(\bv;\boldsymbol\theta) = \boldsymbol\theta^\T \bv$). Alternatively, semiparametric estimators could be developed under the restriction $\gamma_k(\bv)=\gamma_k(\bv;\boldsymbol\theta)$; we leave this to future work. Lastly, another alternative, which is simple and easy to implement, would be to standardize by the marginal standard deviation $\sd(Y_k^0)$ instead of the conditional form  $\sd(Y_k^0 \mid \bV)$; this could be justified merely as an alternative standardization or via the assumption that $\sd(Y_k^0 \mid \bV)=\sd(Y_k^0)$.

\subsection{Weighted Average Summary Measures}
\label{sub:weight}

So far we have discussed robust estimation of outcome-specific effects (scaled for comparability) in Sections \ref{sub:avg}, \ref{sub:quantile}, and \ref{sub:effectmod}, as well as testing of homogeneous effect hypotheses in Section \ref{sub:test}. In many studies it may also be useful to report a summary measure of treatment effect across outcomes. For this purpose we propose weighted average effects of the form
\begin{equation} \label{eq:weight}
\psi^* = \sum_{k=1}^K \E\left\{ w_k(\bV) \frac{ \E(Y_k^1 - Y_k^0 \mid \bV) }{ \sd(Y_k^0 \mid \bV) } \right\} = \sum_{k=1}^K \E\left\{ w_k(\bV) \gamma_k(\bV) \right\} ,
\end{equation}
where $w_k(\bv)$ is an arbitrary user-specified function that indicates how much the summary should be weighted towards outcome $k$ and strata $\bV=\bv$. For example, to weight outcomes equally and to weight strata according to the marginal distribution of $\bV$, one could use $w_k(\bv)=1$; if some outcomes or strata were (a priori) more important, weights could be adjusted accordingly.

An interesting example where weighted average summary measures like $\psi^*$ are useful is as follows. In \textcite{kangovi2017randomized}, a follow-up study to \textcite{kangovi2014patient}, patients were asked (prior to randomization) which of $K$ outcomes they would like to focus on for improvement, so that $V \in \{1,...,K\}$ indicates this choice. Investigators were interested in the overall average effect of treatment $A$ on patients' selected outcomes, which is a special case of the weighted summary in \eqref{eq:weight}.  Specifically, by selecting $w_k(V)=\one(V=k)$ the above summary measure reduces to
$$ \sum_{k=1}^K \Pb(V=k)  \left\{ \frac{ \E(Y_k^1 - Y_k^0 \mid V=k) }{ \sd(Y_k^0 \mid V=k) } \right\}  . $$
This is a weighted average of the scaled effects $\gamma_k(k)$ on selected outcomes (among patients selecting these outcomes), where the weights equal the proportion of patients choosing to focus on each outcome. It should be noted, of course, that these summary measures can provide an obscured view of the multivariate effects $\gamma_k(\bv)$, $k=1, ..., K$, when they are heterogeneous; thus in practice such summary measures should be presented alongside estimates of outcome-specific effects. 

The efficient influence function for the parameter $\psi^*$ in \eqref{eq:weight} is given by
\begin{equation} \label{eq:weight_eif}
\sum_{k=1}^K w_k(\bV) \left\{ \varphi_k^{(\bv)}(\bZ) + \gamma_k(\bV) \right\} - \psi^*
\end{equation}
where $\varphi_k^{(\bv)}(\bZ) $ is the efficient influence function for the $\bv$-specific effect $\gamma_k(\bv)$ given by
\begin{align*}
& \frac{\one(\bV=\bv)}{\Pb(\bV=\bv)} \left( \frac{\phi_{1k}(\bZ) - \phi_{0k}(\bZ)}{\sd(Y_k^0 \mid \bV)}  - \gamma_k(\bv) \left[ \frac{ \phi_{2k}(\bZ) + \E\{(Y_k^0)^2 \mid \bV \} - 2 \E(Y_k^0 \mid \bV) \phi_{0k}(\bZ) }{2 \ \sd(Y_k^0 \mid \bV)^2} \right] \right).
\end{align*}
Therefore, as in previous sections, Wald-type confidence intervals for $\psi^*$ can be constructed by estimating the values of the influence function in \eqref{eq:weight_eif} and using the corresponding empirical standard error.

\section{Simulation Study}


To illustrate some of our proposed methods and explore finite-sample performance, we simulated data with $K=4$ outcomes from the following model:
\begin{equation*}
\begin{gathered}
\bX \sim N(0,I_4) , \\
A \mid \bX \sim \text{Bernoulli}\{ \pi(\bx) \} , \\ 
\pi(\bx) = \text{expit}\{(2x_1 -4 x_2 + 2 x_3 - x_4)/4\} , \\
Y_k \mid \bX, A \sim N\{ \mu_k(\bx,a) , k^2 \}, \\
\mu_k(\bx,a) = k \sum_{j \neq k} (-1)^{j+k-1} x_j + 2 (k-\lambda) a   .
\end{gathered}
\end{equation*}
For the main setting we consider (where $\lambda=2$), the above model gives true values of $\E(Y_k^1 - Y_k^0) = 2 (k-2)$ and $\sd(Y_k^0) =  2k$, so that $\psi_k = 1-2/k$, i.e.,  
$$\boldsymbol\psi = (-1, 0, 1/3, 1/2)^\T . $$ 
The above model also implies that that $\eta_k(\bx,a)$ follows a linear model that is quadratic in the covariates and includes all two-way interactions.

To analyze the above simulated data, we considered estimation of the scaled effect parameter $\boldsymbol\psi$, as well as testing of the homogeneous effects hypothesis discussed in Section 3.3. Here we implemented the proposed estimating equation version of our estimator $\boldsymbol{\hat\psi}$, but also give results for the TMLE version in Section 7 of the Supplementary Materials (with R code for both given in Section 6). This estimator depends on estimates of the nuisance functions $(\pi,\mu,\eta)$, which we constructed using correctly specified parametric models. We used parametric models to ease computation and focus ideas, but in practice we suggest using more flexible methods to minimize risk of model misspecification. To assess potential impacts of such model misspecification in our simulation setup, we fit parametric models with transformed versions of the covariates, using the same transformations as \textcite{kang2007demystifying}. Confidence intervals were constructed using the closed-form influence function-based approach, which is technically only valid under correct modeling of $\pi$; in practice the bootstrap could be used if misspecification of $\pi$ is possible. Results for estimating $\boldsymbol\psi$ are given in Table 1 (with RMSE scaled by $\sqrt{n}$ for easier interpretation).

\begin{table}[h!]
\caption{Results for estimating $\boldsymbol\psi$ across 1000 simulations.}
\label{t:one}
\begin{center}
\begin{tabular}{ll rccc rccc }
\hline
\multicolumn{2}{c}{Correct Model} & \multicolumn{4}{c}{$n=200$} & \multicolumn{4}{c}{$n=1000$}  \\
\multicolumn{2}{c}{\& Parameter} & Bias & SE & RMSE & Cov & Bias & SE & RMSE & Cov  \\ 
\hline
Both 	& $\psi_1$ & -0.02 & 0.14 & 2.03 & 90.2\%
	& -0.00 & 0.06 & 1.97 & 93.5\% \\
	& $\psi_2$ & 0.00 & 0.09 & 1.32 & 93.0\%
	& 0.00 & 0.04 & 1.25 & 95.6\% \\
	& $\psi_3$ & -0.00 & 0.09 & 1.27 & 94.9\%
	& -0.00 & 0.04 & 1.21 & 94.3\% \\
	& $\psi_4$ & 0.02 & 0.11 & 1.54 & 92.8\%
	& 0.00 & 0.05 & 1.52 & 93.9\% \\ 
& & & & & & & & & \\
Trt & $\psi_1$ & -0.05 & 0.16 & 2.67 & 93.7\%
	& -0.01 & 0.08 & 2.63 & 96.8\% \\
	& $\psi_2$ & 0.02 & 0.14 & 1.96 & 97.5\%
	& 0.00 & 0.06 & 1.88 & 98.3\%  \\
	& $\psi_3$ & -0.01 & 0.09 & 1.29 & 93.8\%
	& -0.00 & 0.04 & 1.32 & 95.6\% \\
	& $\psi_4$ & 0.03 & 0.16 & 2.54 & 95.9\%
	& 0.01 & 0.07 & 2.35 & 97.2\% \\
& & & & & & & & & \\
Out & $\psi_1$ & -0.02 & 0.14 & 1.97 & 88.1\%
	& -0.00 & 0.06 & 1.74 & 93.0\% \\
	& $\psi_2$ & 0.01 & 0.09 & 1.34 & 93.6\%
	& 0.00 & 0.04 & 1.23 & 95.3\% \\
	& $\psi_3$ & 0.00 & 0.09 & 1.29 & 93.3\%
	& -0.00 & 0.04 & 1.16 & 95.0\% \\
	& $\psi_4$ & 0.01 & 0.11 & 1.60 & 89.2\%
	& 0.01 & 0.05 & 1.48 & 91.8\% \\
& & & & & & & & & \\
None & $\psi_1$ & -0.31 & 0.20 & 5.24 & 56.5\%
	& -0.30 & 0.09 & 9.90 & 5.8\% \\
	& $\psi_2$ & 0.22 & 0.15 & 3.79 & 59.2\%
	& 0.23 & 0.07 & 7.56 & 6.6\% \\
	& $\psi_3$ & -0.04 & 0.09 & 1.43 & 92.2\%
	& -0.03 & 0.07 & 2.59 & 87.6\% \\
	& $\psi_4$ & 0.36 & 0.21 & 5.93 & 35.9\%
	& 0.37 & 0.09 & 11.98 & 1.7\% \\
\hline
\end{tabular}
\end{center}
\end{table}

The simulations results reflect what is expected based on theory. In particular, the scaled effect $\boldsymbol\psi$ was estimated with small bias whenever either $\pi$ or $(\mu,\eta)$ were correctly modeled, indicating the double robustness of our approach. Of course, if all nuisance estimators are misspecified, no method can promise small bias. Finite-sample biases under correct specification of either $\pi$ or $(\mu,\eta)$ were quite small even when $n=200$, and for $n=1000$ they were almost always zero after rounding (i.e., less than $0.0005$). When all nuisance estimators were correctly modeled, coverage was very close to the nominal 95\% level, especially for the $n=1000$ case. Under misspecification of either $\pi$ or $(\mu,\eta)$, coverage was usually close to 95\% even without any theoretical guarantees, while (as expected) coverage was poor under complete misspecification. In cases where one of $\pi$ or $(\mu,\eta)$ was incorrectly modeled, we expect the bootstrap to provide improved performance. Results for the TMLE estimator were similar (see Section 7 of the Supplementary Materials).

For the hypothesis testing portion of the simulation study, we used the proposed test statistic $T_n$ based on the estimators $\boldsymbol{\hat\psi}$ described earlier, with the covariance estimator $\boldsymbol{\hat\Sigma}$ proposed in Section 3.3 (the estimated covariance of the estimated influence function values). We implemented the approach under the null setting $\lambda=0$ to assess type I error (when $\lambda=2$ and the homogeneous effect hypothesis fails to hold, our approach gave 100\% power in all settings). As before, our theory only guarantees correct error control under correct model specification for both $\mu$ and $(\pi,\eta)$ when using the closed-form variance estimator.

\begin{table}[h!]
\caption{Results for testing homogeneity across 1000 simulations.}
\label{t:two}
\begin{center}
\begin{tabular}{lrrrr}
\hline
Correct & \multicolumn{4}{c}{Type I Error} \\
Model  & $n=200$ & $n=500$ & $n=1000$ & $n=5000$ \\
\hline
Both & 11.4\% & 8.4\% & 5.8\% & 5.5\% \\
Trt  & 9.0\% & 6.4\% & 5.6\% & 3.8\% \\
Out  & 13.5\% & 9.3\% & 7.7\% & 7.5\% \\
None & 55.7\% & 91.2\% & 99.2\% & 100\% \\
\hline
\end{tabular}
\end{center}
\end{table}

The simulations show that our test approximately controls type I error at nominal rates in large samples, if either working model is correct. There is some anticonservative bias for smaller sample sizes, as is often the case for generalized Wald tests \autocite{boos2013essential}. We expect this could be ameliorated by using the bootstrap instead of a closed-form variance estimator. Although the closed-form variance estimator does not guarantee error control under misspecification, the type I error was relatively close to 5\% as long as one of $\pi$ or $(\mu,\eta)$ was correctly modeled. Again TMLE results are given in Section 7 of the Supplementary Materials; in this setting the TMLE version of our estimator gave inflated type I error relative to the estimating equation results presented in the main text.

\section{Application}
\label{s:app}

Here we apply our proposed methods to a recent trial \autocite{kangovi2017randomized} studying the effects of a (randomized) community health worker intervention on four chronic disease-related outcomes: cigarettes per day (CPD), systolic blood pressure (SBP), HbA1c, and body mass index (BMI). At baseline each subject was asked which of the outcomes he or she would prefer to focus on, and the primary goal of the study was to learn whether the intervention affected patients' focus outcomes. Each outcome was measured in terms of change between study enrollment and 6-month follow-up, so negative values indicate improvement. \textcite{kangovi2017randomized} give full details of the study population and design.  

In our analysis we adjust for baseline outcomes, age, gender, and the selected focus outcome, and we used the cross validation-based Super Learner \autocite{van2011targeted} to combine parametric models (linear for the outcome, logistic for the treatment), generalized additive models, and random forests. Note however that here covariate adjustment is only for the purposes of increasing efficiency, since the intervention was completely randomized (i.e., our propensity score model is guaranteed to be correctly specified). We estimate the scaled effects from Section 3.1, test effect homogeneity as in Section 3.3, estimate how effects vary with $V$ = selected outcomes as in Section 3.4, and finally estimate the average effect on selected outcomes via the summary measure approach from Section 3.5.

\begin{figure}[h!]
\caption{Estimates of scaled effects $\psi_k$ (left) and effects on focus outcomes $\gamma_k(k)$ (right), with pointwise 95\% CIs.  Also shown is the p-value for testing effect homogeneity (left) and the estimated summary effect on selected outcomes with 95\% CI (right). \label{fig:res}}
\begin{center}
\includegraphics[width=\textwidth]{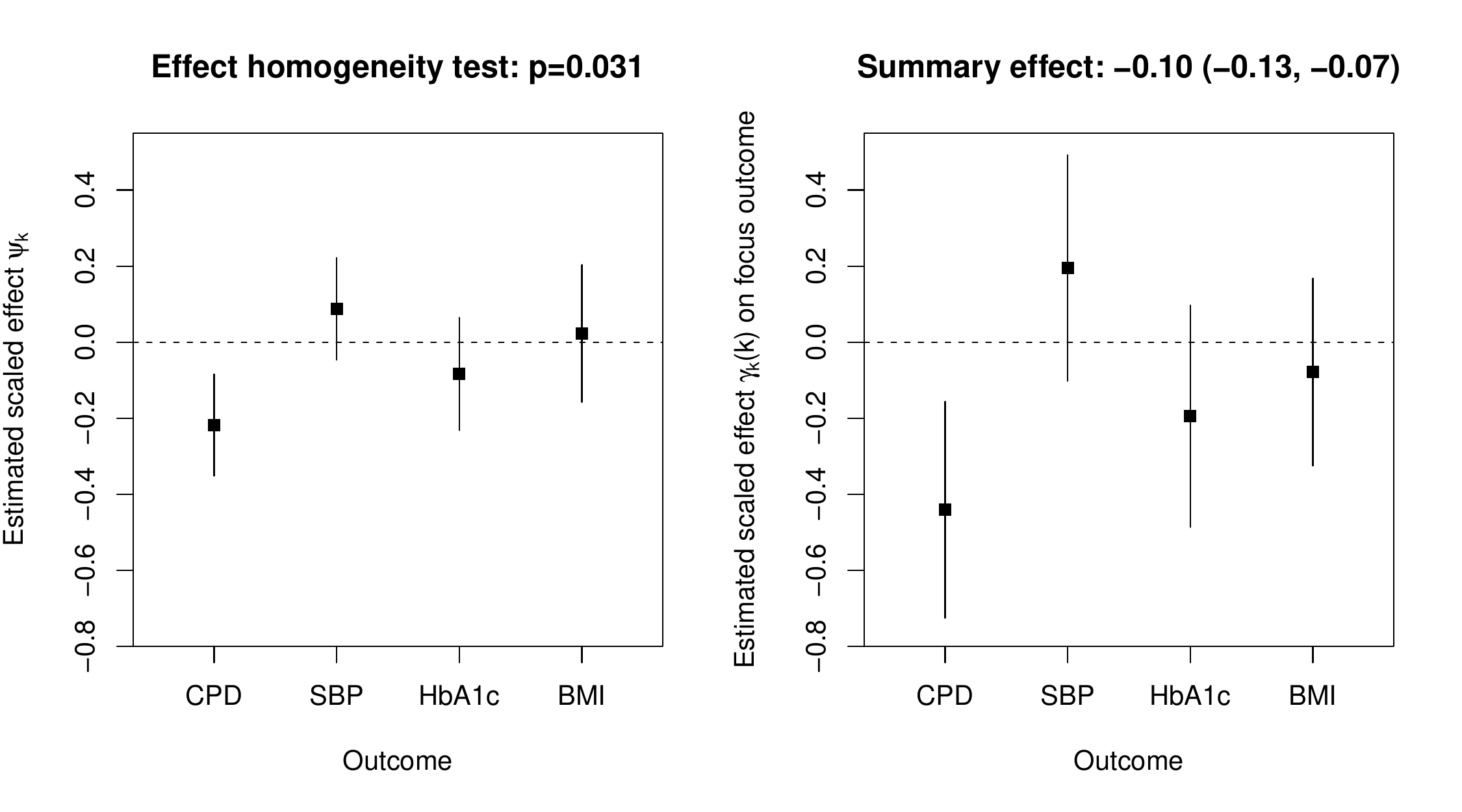}
\end{center}
\end{figure}

Results are displayed in Figure~\ref{fig:res}. The scaled effect estimates (with non-simultaneous 95\% confidence intervals) were -0.22 (-0.36, -0.08) for CPD, 0.09 (-0.04, 0.22) for SBP, -0.08 (-0.22, 0.05) for HbA1c, and 0.02 (-0.15, 0.19) for BMI. These results indicate that the intervention was effective in reducing cigarettes smoked per day, yielding a decrease of roughly a fifth of a standard deviation (of pre-post differences). The null hypothesis of zero scaled effect for CPD was in fact rejected at level 0.05 after Bonferroni correction for multiple testing (since $p=0.001 < 0.05/4$). There was some effect homogeneity, as indicated in the left panel of Figure~\ref{fig:res} and by the fact that the homogeneous effects test from Section 3.3 was rejected ($p=0.031$). Hence we have evidence that the intervention affects different outcomes differently. Finally we also see that effects were generally stronger for outcomes that patients selected to focus on. However it is still only the CPD outcome for which we can reject a non-zero effect at the 0.05 level ($p=0.003$). The estimated average effect on selected outcomes was a tenth of a standard deviation, and significantly different from zero. However it is clear that much of this effect comes from the effects on smoking.

\section{Discussion}
\label{s:discuss}

In this paper we proposed flexible methods for estimating and testing effects in studies with multiple outcomes. We developed nonparametric doubly robust methods for estimating effects scaled using mean-variance and interquartile-range standardizations, including effect modification and weighted summary measures, and constructed a test of effect homogeneity. We expect our work to be important for both randomized trials and observational studies, and feel our distribution-free results fill an important gap in the literature.

There are a number of important future directions to this research. In an upcoming paper we will apply the methods to explore effects of a community health worker intervention in a patient-centered study that allows patients to choose to focus on certain outcomes rather than others, as briefly described in Section 3.5. It will also be useful to develop tests of other hypotheses beyond homogeneity (e.g., tests of no effect), to more thoroughly explore the implementation and performance of the bootstrap or other methods for weakening  assumptions for valid inference, and to consider an asymptotic regime in which $K=K_n$ increases with sample size (which may more accurately represent studies with many outcomes).

\stepcounter{section}
\printbibliography[title={\thesection \ \ \ References}]

\pagebreak
\setcounter{page}{1}

\vspace*{.2in}
\begin{center}
\LARGE \textbf{Supplementary Materials for \\ ``Estimating scaled treatment effects \\ with multiple outcomes''}
\end{center}

\vspace*{.5in}

\normalsize

\setcounter{section}{0}
\section{Proof of double robustness of $\hat\psi_k$}

Using iterated expectation, it is straightforward to show that $\phi_{ak}$ and $\phi_{2k}$ are doubly robust in the sense that 
\begin{align*}
\E\{\phi_{ak}(\bZ;\overline\pi,\overline\mu)\} &= \E(Y_k^a) \\
\E\{\phi_{2k}(\bZ;\overline\pi,\overline\eta)\} &= \E\{(Y_k^0)^2\}
\end{align*}
as long as either $\overline\pi=\pi$ or $(\overline\mu,\overline\eta)=(\mu,\eta)$, not necessarily both. \\

Thus under standard Glivenko-Cantelli regularity conditions on the estimators $(\hat\pi,\hat\mu,\hat\eta)$ and their limits $(\overline\pi,\overline\mu,\overline\eta)$, as long as either $\overline\pi=\pi$ or $(\overline\mu,\overline\eta)=(\mu,\eta)$, then we have
\begin{align*}
\Pn\Big\{ \phi_{ak}(\bZ;\hat\pi,\hat\mu) \Big\} &\inprob \E(Y_k^a) \\
\Pn\Big\{ \phi_{2k}(\bZ;\hat\pi,\hat\eta) \Big\} &\inprob \E\{(Y_k^0)^2\} .
\end{align*}
Therefore, by the continuous mapping theorem (and if $\sd(Y_k^0)>0$ so $\psi_k$ is well-defined), we have 
$$ \hat\psi_k \inprob \frac{ \E(Y_k^1) - \E(Y_k^0) }{ \sqrt{ \E\{(Y_k^0)^2\} - \E(Y_k^0)^2 } } = \psi_k $$
as long as either $\hat\pi$ or $(\hat\mu,\hat\eta)$ converge to the truth, and so $\hat\psi_k$ is doubly robust.

Double robustness can also be seen via the efficient influence function, since 
\begin{align*}
\E &\left( \frac{\phi_{1k} - \phi_{0k} }{ \sd(Y_k^0) } - \psi_k \left[ \frac{ \phi_{2k} + \E\{(Y_k^0)^2\} - 2\E(Y_k^0) \phi_{0k} }{ 2 \sd(Y_k^0)^2 } \right] \right) = 0 \\
\hspace{.4in} &\implies \psi_k = \frac{2\sd(Y_k^0) \E(\phi_{1k}-\phi_{0k})}{\E[\phi_{2k} + \E\{(Y_k^0)^2\} - 2\E(Y_k^0) \phi_{0k}] } = \frac{ \E(Y_k^1 - Y_k^0)}{ \sd(Y_k^0)} 
\end{align*}
and the last step follows as long as either $\overline\pi=\pi$ or $(\overline\mu,\overline\eta)=(\mu,\eta)$.

\section{Proof of asymptotic normality}

Let $\boldsymbol{\beta}_k = (\beta_{0k},\beta_{1k},\beta_{2k})^\T$ with
$$  \beta_{0k} = \E(Y^0_k) \ , \ \beta_{1k} = \E(Y^1_k) \ , \ \beta_{2k} = \E\{(Y^0_k)^2\} $$
and define the corresponding estimator $\boldsymbol{\hat\beta}_k = (\hat\beta_{0k},\hat\beta_{1k},\hat\beta_{2k})^\T$ for
$$ \hat\beta_{0k} = \Pn\{ \phi_{0k}(\bZ;\hat\pi,\hat\mu) \}  \ , \ \hat\beta_{1k} = \Pn\{ \phi_{1k}(\bZ;\hat\pi,\hat\mu) \}  \ , \ \hat\beta_{2k} = \Pn\{ \phi_{2k}(\bZ;\hat\pi,\hat\eta) \}  . $$
Let $\boldsymbol\phi_k=(\phi_{0k},\phi_{1k},\phi_{2k})^\T$, and suppose that 
\begin{enumerate}
\item $(\overline\pi,\overline\mu,\overline\eta)=(\pi,\mu,\eta)$, 
\item $||\hat\pi-\pi ||=o_\Pb(n^{-1/4})$ and $(||\hat\mu-\mu|| + ||\hat\eta - \eta ||)=o_\Pb(n^{-1/4})$, 
\item $(\overline\pi,\overline\mu,\overline\eta)$ and $(\hat\pi,\hat\mu,\hat\eta)$ fall in Donsker classes.
\end{enumerate} 
Then by for example Theorem 5.31 for Z-estimators from \textcite{van2000asymptotic} we have
$$ \boldsymbol{\hat\beta}_k - \boldsymbol{\beta}_k = \Pn \Big\{ \boldsymbol\phi_k(\bZ;\pi,\mu,\eta) - \boldsymbol\beta_k \Big\} + o_\Pb(1/\sqrt{n}) . $$
Now since  $\hat\psi_k = g(\boldsymbol{\hat\beta}_k) = ( \hat\beta_{1k}-\hat\beta_{0k} ) / { \sqrt{ \hat\beta_{2k} - \hat\beta_{0k}^2} }$,  an application of the delta method (with detailed calculations given in the next section of these Supplementary Materials) yields
\begin{align*}
 \hat\psi_k - \psi_k &= \Pn\Big[  (\nabla g) \Big\{ \boldsymbol\phi_k(\bZ;\pi,\mu,\eta) - \boldsymbol\beta_k \Big\} \Big] + o_\Pb(1/\sqrt{n}) \\
 &= \Pn \left[ \frac{\phi_{1k} - \phi_{0k} }{ \sqrt{\beta_{2k}-\beta_{0k}^2} } - \psi_k \left\{ \frac{ \phi_{2k} + \beta_{2k} - 2 \beta_{0k} \phi_{0k} }{ 2 (\beta_{2k}-\beta_{0k}^2) } \right\}  \right] + o_\Pb(1/\sqrt{n}) . 
 \end{align*}
Asymptotic normality then follows immediately from the central limit theorem.

\section{Derivation of efficient influence function}

That the efficient influence function is
$$ \varphi_k = \frac{\phi_{1k} - \phi_{0k} }{ \sd(Y_k^0) } - \psi_k \left[ \frac{ \phi_{2k} + \E\{(Y_k^0)^2\} - 2\E(Y_k^0) \phi_{0k} }{ 2 \sd(Y_k^0)^2 } \right] $$ 
follows from the fact that $(\phi_{ak}-\beta_{ak})$ and $(\phi_{2k}-\beta_{2k})$ are the efficient influence functions for $\beta_{ak}$ and $\beta_{2k}$, respectively, together with the delta method.

Specifically, for $\psi_k = g(\boldsymbol\beta_k) = ( \beta_{1k}-\beta_{0k} ) / { \sqrt{ \beta_{2k} - \beta_{0k}^2} }$ we have
\begin{align*}
\nabla g &= \left\{ \frac{\partial g(\boldsymbol\beta_k)}{\partial \beta_{0k}} , \frac{\partial g(\boldsymbol\beta_k)}{\partial \beta_{1k}}  , \frac{\partial g(\boldsymbol\beta_k)}{\partial \beta_{2k}}  \right\} \\
&=  \frac{1}{\sqrt{\beta_{2k} - \beta_{0k}^2}}   \left\{ \left( \frac{\beta_{0k} \beta_{1k} - \beta_{2k} }{\beta_{2k}-\beta_{0k}^2 }  \right), 1 , \frac{1}{2} \left( \frac{\beta_{0k} - \beta_{1k} }{\beta_{2k}-\beta_{0k}^2 }  \right) \right\} \\
&= \frac{1}{\sd(Y_k^0)} \Big\{ \psi_k \beta_{0k}/\sd(Y_k^0) -1 , 1, -\psi_k / 2\sd(Y_k^0)^2 \Big\}
\end{align*}
so that, letting $\tau_k=\sd(Y_k^0)$, 
\begin{align*}
(\nabla g) \Big( \boldsymbol\phi_k - \boldsymbol\beta_k \Big) &= \frac{1}{\tau_k} \left[ \left( \frac{\beta_{0k}}{\tau_k} \psi_k - 1\right) (\phi_{0k} - \beta_{0k}) + (\phi_{1k}-\beta_{1k}) - \psi_k (\phi_{2k}-\beta_{2k}) / 2\tau_k \right] \\
&=  \frac{ (\phi_{1k} - \beta_{1k} ) - (\phi_{0k}-\beta_{0k}) }{\tau_k} - \frac{\psi_k}{2 \tau_k^2}\Big\{ (\phi_{2k}-\beta_{2k}) - 2(\phi_{0k}-\beta_{0k}) \beta_{0k} \Big\} \\
&= \left( \frac{ \phi_{1k} - \phi_{0k} }{\tau_k} \right)  - \psi_k - \psi_k \left( \frac{\phi_{2k} - 2\phi_{0k} \beta_{0k} + \beta_{2k}}{2 \tau_k^2} \right) + \psi_k \\
&= \frac{\phi_{1k} - \phi_{0k} }{ \sd(Y_k^0) } - \psi_k \left[ \frac{ \phi_{2k} + \E\{(Y_k^0)^2\} - 2\E(Y_k^0) \phi_{0k} }{ 2 \sd(Y_k^0)^2 } \right]
\end{align*}
as given in the main text.

\section{Efficient influence function (quantile case)}

As discussed by \textcite{diaz2015efficient} the efficient influence function for $\xi_{aq} = F_{ak}^{-1}(q)$ is given by
$$ \phi^q_{ak} = \frac{-1}{\Pb(Y_k=\xi_{aq})} \left[ \frac{\one(A=a) }{ \pi(a \mid \bX)}  \Big\{\one(Y_k \leq \xi_{aq}) - \nu_k(\xi_{aq} \mid \bX,a) \Big\}  + \nu_k(\xi_{aq} \mid \bX,a) - q \right] . $$
Letting $\boldsymbol\xi_k=(\xi_{1,.5},\xi_{0,.5},\xi_{0,.75},\xi_{0,.25})$, then if $\boldsymbol{\hat\xi}_k$ is an efficient estimator (e.g., solving the efficient influence function estimating equation up to order $o_\Pb(1/\sqrt{n})$) we have
\begin{align*}
\boldsymbol{\hat\xi}_k - \boldsymbol\xi_k = \Pn (\boldsymbol\phi_k^q) + o_\Pb(1/\sqrt{n})
\end{align*}
where $\boldsymbol\phi_k^q = (\phi_{1k}^{.5}, \phi_{0k}^{.5}, \phi_{0k}^{.75}, \phi_{0k}^{.25})^\T$. Then by the delta method the efficient influence function for $\psi_k^q = (\xi_{1,.5}-\xi_{0,.5})/(\xi_{0,.75}-\xi_{0,.25})$ is given by
$$ \frac{(\phi_{1k}^{.5} - \phi_{0k}^{.5}) - \psi_k^q (\phi_{0k}^{.75} - \phi_{0k}^{.25})}{ \xi_{0,.75}-\xi_{0,.25}} . $$

\section{Asymptotic distribution of test statistic}

This section relies on the asymptotic linearity result from Section 2. Under the null hypothesis, due to the banded structure of $\mathbf{C}$ we have $\mathbf{C} \boldsymbol\psi = \mathbf{0}$, so that since
$$\boldsymbol{\hat\psi} - \boldsymbol\psi = \Pn( \boldsymbol\varphi) + o_\Pb(1/\sqrt{n})$$ 
it follows that
$$  \mathbf{C} \boldsymbol{\hat\psi} = \Pn( \mathbf{C} \boldsymbol\varphi) + o_\Pb(1/\sqrt{n}) . $$
Therefore $\sqrt{n} \mathbf{C} \boldsymbol{\hat\psi} \indist N\{0,\cov(\mathbf{C} \boldsymbol\varphi)\}$ by the central limit theorem, and similarly for the corresponding quadratic form we have
$$ (\sqrt{n} \mathbf{C} \boldsymbol{\hat\psi})^\T \cov( \mathbf{C} \boldsymbol\varphi)^{-1} (\sqrt{n} \mathbf{C} \boldsymbol{\hat\psi}) = n (\mathbf{C} \boldsymbol{\hat\psi})^\T ( \mathbf{C} \boldsymbol\Sigma \mathbf{C}^\T)^{-1} (\mathbf{C} \boldsymbol{\hat\psi}) \indist \chi^2_{K-1}. $$
By Slutsky's theorem the same result holds when $\boldsymbol{\hat\Sigma} \inprob \boldsymbol\Sigma$ replaces $\boldsymbol\Sigma$.


\section{R code for simulation}

\begin{verbatim}

library(tmle)
expit <- function(x){ exp(x)/(1+exp(x)) }
logit <- function(x){ log(x/(1-x)) }

n <- 1000
nsim <- 500
cor <- "both" # options for correct model(s): both, trt, out, none
nullcase <- F

## note: to implement tmle version replace `Qinit$Q' with `Qstar'

## create matrix to store results
res.names <- c(paste("psi",1:4,sep=""),paste("psi.se",1:4,sep=""),"tn","pval")
res <- data.frame(matrix(NA,nrow=nsim,ncol=length(res.names)))
colnames(res) <- res.names; i <- 1

for (i in 1:nsim){
print(i); flush.console()

## simulate data
x <- matrix( rnorm(4*n) , nrow=n); colnames(x) <- paste("x",1:4,sep="")
pi <- expit((2*x[,1] -4*x[,2] + 2*x[,3] - x[,4])/4)
a <- rbinom(n,1,pi)
kmat <- matrix(rep(1:4,n),nrow=n,byrow=T)
mu <- kmat * cbind( x[,2]-x[,3]+x[,4], x[,1]+x[,3]-x[,4], 
  -x[,1]+x[,2]+x[,4], x[,1]-x[,2]+x[,3] ) + 2*(kmat-2*(!nullcase))*a
y <- mu + matrix(rnorm(4*n,sd=kmat),nrow=n); colnames(y) <- paste("y",1:4,sep="")

## construct covariates
xm <- cbind(exp(x[,1]/2), 10 + x[,2]/(1+exp(x[,1])), 
  (.6+x[,1]*x[,3]/25)^3, (x[,2]+x[,4]+20)^2)
if (cor=="both"){ x <- data.frame(cbind(x,x)) }
if (cor=="out"){ x <- data.frame(cbind(xm,x)) }
if (cor=="trt"){ x <- data.frame(cbind(x,xm)) }
if (cor=="none"){ x <- data.frame(cbind(xm,xm)) }
x <- cbind(x, x[,5:8]^2, x[,5]*x[,6], x[,5]*x[,7], x[,5]*x[,8], 
  x[,6]*x[,7], x[,6]*x[,8], x[,7]*x[,8])
colnames(x) <- c(paste("gx",1:4,sep=""), paste("qx",1:4,sep=""),
  paste("q2x",1:10,sep=""))
  

## estimation/inference for psi
infvals <- NULL; for (j in 1:4){
  ## estimate b1=EY1, obtain inf fn vals
  b1.tmle <- tmle(Y=y[,j],A=NULL,W=x,Delta=a, 
    g.Deltaform=Delta~gx1+gx2+gx3+gx4, Qform=Y~qx1+qx2+qx3+qx4)
  phi1 <- a*(y[,j]-b1.tmle$Qinit$Q[,2]) / b1.tmle$g.Delta$g1W[,1] + 
    b1.tmle$Qinit$Q[,2]; b1 <- mean(phi1)
  ## estimate b0=EY0, obtain inf fn vals  
  b0.tmle <- tmle(Y=y[,j],A=NULL,W=x,Delta=1-a, 
    g.Deltaform=Delta~gx1+gx2+gx3+gx4, Qform=Y~qx1+qx2+qx3+qx4)
  phi0 <- (1-a)*(y[,j]-b0.tmle$Qinit$Q[,2]) / b0.tmle$g.Delta$g1W[,1] + 
    b0.tmle$Qinit$Q[,2]; b0 <- mean(phi0)
  ## estimate b2=E{(Y0)^2}, obtain inf fn vals   
  b2.tmle <- tmle(Y=y[,j]^2,A=NULL,W=x,Delta=1-a, g.Deltaform=Delta~gx1+gx2+gx3+gx4, 
    Qform=Y~q2x1+q2x2+q2x3+q2x4+q2x5+q2x6+q2x7+q2x8+q2x9+q2x10)
  phi2 <- (1-a)*(y[,j]^2-b2.tmle$Qinit$Q[,2]) / b2.tmle$g.Delta$g1W[,1] + 
    b2.tmle$Qinit$Q[,2]; b2 <- mean(phi2) 
  ## estimate scaled effect psi
  res[i,j] <- (b1-b0)/sqrt(b2-b0^2)
  ## get inf fn vals for psi
  infvals <- cbind(infvals, (phi1 - phi0)/sqrt(b2-b0^2) - res[i,j] *
    (phi2 + b2 - 2*b0*phi0)/(2*(b2-b0^2)) ) }
## compute variance based on inf fn vals
res[i,5:8] <- sqrt(diag(cov(infvals))/n)

## test homogeneity hypothesis
cmat <- rbind(c(1,-1,0,0),c(0,1,-1,0),c(0,0,1,-1))
psi <- t(res[i,1:4]); sigma <- cov(infvals)
res$tn[i] <- n * t(cmat %*% psi) %*% solve( cmat %*% sigma %*% t(cmat)) %*% 
  (cmat %*% psi); res$pval[i] <- pchisq(res$tn[i],df=3,lower.tail=F)

}

## summarize simulation results
if (nullcase==F){ psi0 <- c(-1,0,1/3,.5) }; if (nullcase==T){ psi0 <- rep(1,4) }
psimat <- matrix(psi0,nrow=nsim,ncol=4,byrow=T)
(resmat <- data.frame(
  bias=apply(res[,1:4],2,mean,na.rm=T)-psi0, se=apply(res[,1:4],2,sd,na.rm=T),
  med.se=apply(res[,5:8],2,median,na.rm=T), 
  rmse=sqrt(n*apply((res[,1:4]-psimat)^2,2,mean,na.rm=T)), 
  cov=apply((res[,1:4]-1.96*res[,5:8]<psimat) & 
    (res[,1:4]+1.96*res[,5:8]>psimat),2,mean, na.rm=T )) )
mean(res$pval<=0.05, na.rm=T)
\end{verbatim}

\section{TMLE simulation results}

\begin{table}[h!]
\caption{Results for estimating $\boldsymbol\psi$ via TMLE across 500 simulations.}
\label{t:one}
\begin{center}
\begin{tabular}{ll rccc rccc }
\hline
\multicolumn{2}{c}{Correct Model} & \multicolumn{4}{c}{$n=200$} & \multicolumn{4}{c}{$n=1000$}  \\
\multicolumn{2}{c}{\& Parameter} & Bias & SE & RMSE & Cov & Bias & SE & RMSE & Cov  \\ 
\hline
Both 	& $\psi_1$ & -0.02 & 0.14 & 1.98 & 87.8\%
	& -0.00 & 0.06 & 1.97 & 94.0\% \\
	& $\psi_2$ & 0.01 & 0.09 & 1.31 & 92.2\%
	& 0.00 & 0.04 & 1.27 & 95.0\% \\
	& $\psi_3$ & -0.01 & 0.09 & 1.22 & 93.8\%
	& -0.00 & 0.04 & 1.26 & 93.8\% \\
	& $\psi_4$ & 0.02 & 0.11 & 1.51 & 92.4\%
	& 0.00 & 0.05 & 1.55 & 94.6\% \\ 
& & & & & & & & & \\
Trt & $\psi_1$ & -0.05 & 0.16 & 2.42 & 91.6\%
	& -0.01 & 0.07 & 2.29 & 96.8\% \\
	& $\psi_2$ & 0.02 & 0.11 & 1.61 & 97.6\%
	& 0.00 & 0.05 & 1.64 & 97.0\%  \\
	& $\psi_3$ & -0.01 & 0.09 & 1.26 & 93.8\%
	& -0.00 & 0.04 & 1.32 & 94.2\% \\
	& $\psi_4$ & 0.04 & 0.12 & 1.87 & 96.6\%
	& 0.01 & 0.05 & 1.71 & 97.8\% \\
& & & & & & & & & \\
Out & $\psi_1$ & -0.02 & 0.12 & 1.77 & 88.6\%
	& -0.00 & 0.05 & 1.69 & 92.6\% \\
	& $\psi_2$ & 0.01 & 0.09 & 1.26 & 91.8\%
	& 0.00 & 0.04 & 1.24 & 93.2\% \\
	& $\psi_3$ & -0.00 & 0.09 & 1.24 & 93.4\%
	& -0.00 & 0.04 & 1.21 & 93.6\% \\
	& $\psi_4$ & 0.01 & 0.09 & 1.33 & 92.4\%
	& 0.00 & 0.04 & 1.34 & 94.4\% \\
& & & & & & & & & \\
None & $\psi_1$ & -0.34 & 0.20 & 5.50 & 49.8\%
	& -0.31 & 0.09 & 10.25 & 5.0\% \\
	& $\psi_2$ & 0.24 & 0.15 & 4.00 & 51.6\%
	& 0.23 & 0.07 & 7.67 & 5.4\% \\
	& $\psi_3$ & -0.03 & 0.09 & 1.35 & 91.6\%
	& -0.03 & 0.04 & 1.52 & 90.8\% \\
	& $\psi_4$ & 0.39 & 0.18 & 6.11 & 30.8\%
	& 0.38 & 0.08 & 12.38 & 0.8\% \\
\hline
\end{tabular}
\end{center}
\end{table}

\begin{table}[h!]
\caption{Results for testing homogeneity via TMLE across 1000 simulations.}
\label{t:two}
\begin{center}
\begin{tabular}{lrrrr}
\hline
Correct & \multicolumn{4}{c}{Type I Error} \\
Model & $n=200$ & $n=500$ & $n=1000$ & $n=5000$ \\
\hline
Both  & 12.9\% & 9.1\% & 6.6\% & 6.1\% \\
Trt  & 10.3\% & 7.2\% & 6.5\% & 3.2\% \\
Out  & 13.6\% & 10.2\% & 7.9\% & 7.6\% \\
None  & 58.2\% & 92.9\% & 99.5\% & 100.0\% \\
\hline
\end{tabular}
\end{center}
\end{table}

\end{document}